\documentclass[11pt]{article}
\usepackage{amsfonts%,showkeys
}
\newcommand{\be}{\begin{equation}}
\newcommand{\ee}{\end{equation}}
\newcommand{\beqn}{\begin{eqnarray}}
\newcommand{\eeqn}{\end{eqnarray}}
\newcommand{\Sch}{Schr\"odinger }
\newcommand{\Vmax}{V{\mbox{\scriptsize max}}}
\newcommand{\Vmin}{V{\mbox{\scriptsize min}}}
\newcommand{\tg}{\tilde{g}}
\newcommand{\tf}{\tilde{f}}
\newcommand{\sth}{{h}}
\newcommand{\dth}{\smash{{\tilde{h}}}}
\newcommand{\tth}{\smash{{\tilde{\tilde{h}}}}}
\newcommand{\vf}{\mbox{\it v}}
\newcommand{\fg}{\mbox{$f\!g$}}
\newcommand{\gh}{\mbox{$g\!h$}}
\newcommand{\fh}{\mbox{$f\!h$}}
\newcommand{\sn}{\mbox{sn}}
\newcommand{\cn}{\mbox{cn}}
\newcommand{\dn}{\mbox{dn}}

\begin{document}
\begin{center}

{\large {\bf A General Approach of Quasi-Exactly Solvable
 \Sch Equations}}\\ \vspace{2cm}
 N. DEBERGH\footnote{Email: Nathalie.Debergh@ulg.ac.be}(a),
J. NDIMUBANDI\footnote{Email: jndimubandi@yahoo.fr}(b), \\
and B.
VAN DEN BOSSCHE\footnote{Email: bvandenbossche@ulg.ac.be}(a)\\

\vspace{2cm}

(a){\it Fundamental Theoretical Physics,
Institute of Physics (B5),
University of Li\`ege,
B-4000 LIEGE  (Belgium)}\\ \vspace{1cm}

(b){\it University of Burundi, Department of Mathematics,
 P.O. Box 2700, BUJUMBURA (Burundi)} \\
%\vspace{0.2in}
\end{center}
%\vspace{0.5in}
\begin{abstract}
We construct a general algorithm generating the analytic eigenfunctions
as well as eigenvalues of one-dimensional stationary \Sch Hamiltonians.
Both exact and quasi-exact Hamiltonians enter our formalism but we focus
on quasi-exact interactions for which no such general approach has been
considered before. In particular we concentrate on a generalized sextic
oscillator but also on the Lam\'e and the screened Coulomb potentials.
\end{abstract}

%\vfill
%{\bf number of pages : 39, no figure, no table}

\newpage

%Running title : Solving Q.E.S. equations\\

%\vspace{5cm}

%Contact author : \\
%Dr. N. Debergh \\
%Fundamental Theoretical Physics\\
%Institute of Physics (B5)\\
%University of Li\`ege\\
%B-4000 LIEGE  (Belgium)\\
%Email: Nathalie.Debergh@ulg.ac.be\\
%Tel : +32-4-366.36.40\\
%Fax : +32-4-366.36.72

%\newpage

\section{Introduction}
\label{Section1}
 Quasi-exactly solvable (Q.E.S.) \Sch equations
 nowadays attract much attention. They are equations for
 which a finite number of solutions can be analytically
 determined. In a sense, they constitute the intermediate
 step between exactly solvable equations (associated with
 potentials such as the harmonic oscillator one, the Coulomb one,...)
 whose all analytic solutions can be obtained, and the analytically unsolvable
 ones requiring a numerical treatment. These Q.E.S. equations have been
 studied following two different points of view.

-- First, the number of solutions
 can be related \cite{ref1,ref2,ref3,ref4}
to the dimension of the irreducible representations of a Lie algebra,
namely $sl(2,R)$. In that case the corresponding \Sch Hamiltonian
can be written as  linear and quadratic combinations of the $sl(2,R)$
generators, these ones preserving the finite-dimensional space of the
solutions. Due to the preservation of this space, the associated
 time-independent differential \Sch equation actually reduces
to an algebraic one.

-- Second, this reduction from differential to algebraic is maintained but
the number of known solutions  (essentially two \cite{ref5,ref6} or three
\cite{ref7})
is fixed at the start. Generally speaking this second point of view is
not subtended by any Lie algebra except in the case of the two
solutions~\cite{ref8}.
Instead, it is relevant of supersymmetric quantum mechanics \cite{ref9}.

The main purpose of this paper is to propose
a general approach unifying these two points of view.
Our approach,
which is also algebraic and, for the moment, restricted to the real
 line $]-\infty,\infty[$, or half line $]0,\infty[$,
is well-suited for all potentials
leading to a solution given by an overall function modulated by a
polynomial. To our knowledge, it reproduces all
the known exact cases which can be expressed in terms of polynomials,
and most, if not all, of Q.E.S. models.
%(We are still investigating if all
%the Q.E.S. models with periodic potentials constructed from
%Jacobi Elliptic functions can be included in our formalism).
For example, for the exact
cases, it reproduces solutions of the harmonic oscillator,
the Morse and P\"oschl-Teller
potentials, solutions of the Lam\'e equation
\cite{yves95}, the even potentials of the class
described recently
by Bender and Wang in Ref. \cite{bw01}. On the contrary, since
the odd potentials of the latter
class cannot be written in term of polynomials
(the solution is written as a non-terminating confluent hypergeometric
function), they cannot be dealt with.
As examples of Q.E.S. \Sch equations, we can quote all the potentials
given by Turbiner in \cite{ref3}, a variation on the
Lam\'e equation as described in \cite{khare01},
as well as  the Tkachuk potential of
Ref.~\cite{ref5}. The list is non-exhaustive.

The paper is organized as follows:
In Section~\ref{Section2}, we explain the  formalism of our method in
a simplifying limit regarding the expansion basis.
In Section~\ref{Section3}, we develop the most general algorithm
based on a single basis function, and restricted to finite sums.
In Section~\ref{Section4}, we show the relevance of the method by
investigating several
well-known Q.E.S. equations, of both types quoted above. We  show
in Section~\ref{Section5} that it can be applied to yet unused potential by
considering a modification of the sextic oscillator potential.
%{\bf In Section~\ref{Section5bis}, we review some group theoretical bla-bla
% ... }.
Finally, we draw our conclusions in Section~\ref{Section6}.

\section{A Q.E.S. general approach}
\label{Section2}
As stated in the Introduction,
we concentrate on the one-dimensional and time-independent \Sch equation
\be
\left[-\frac{d^2}{dx^2} + V(x)\right]\psi_N (x)
= E_N \psi_N(x), \; N=0, 1, 2, ...
\label{Eq1}
\ee
for which a finite number ($=n+1$) of solutions $\psi_N(x)$
%($N$ fixed)
can be
 determined.
The procedure we shall explain here is not the most general one, which can
be found in Section~\ref{Section3}. We proceed so because
we want to give here more details relevant to
understand the philosophy of our approach.
%Such details would be
%too cumbersome in the full algorithm.
In order to cover the Lie algebraic and non-Lie algebraic Q.E.S.
 equations we write the potential $V(x)$ as
\be
V(x) = \sum_{k=-2}^{\Vmax} V_k [f(x)]^k,
\label{Eq2}
\ee
$\Vmax$ being a positive integer, while the coefficients
$V_k$ and the function
$f(x)$ are, for the moment, arbitrary. Making the sum begin with the
index $k=-2$ allows to take care of the centrifugal barrier in the radial
equation of a 3-dimensional problem. We shall see in Eqs.~(\ref{Eq10})
and~(\ref{Eq12}) that the minimal value of $k$ has anyway to be $k=-2$
in order to factor out a common power of $f(x)$.

The eigenfunctions $\psi_N(x)$ are assumed to be
\be
\psi_N(x)= g(x) [f(x)]^{\lambda_N} \sum_{m=0}^N c_m^{(N)} [h(x)]^m.
\label{Eq3}
\ee
The function $g(x)$ plays the role of a weight factor,
$[f(x)]^{\lambda_N}$ is introduced due to eventual singularities
in~(\ref{Eq2}), and the quantities $c_m^{(N)}$
are expansion coefficients on the basis $h(x)$.
We  introduce the notation
\be
\psi_N(x)=\sum_{m=0}^N  c_m^{(N)} \psi_m(x),
\label{Eq4}
\ee
with
\be
\psi_m(x) = g(x) [f(x)]^{\lambda_N} [h(x)]^m.
\label{Eq5}
\ee
We then have
\beqn
&&\left[-\frac{d^2}{dx^2} + V(x)\right] \psi_m (x)
= \bigg\{-\frac{g''(x)}{g(x)}-\lambda_N \frac{f''(x)}{f(x)}
-m\frac{h''(x)}{h(x)}
\nonumber \\
&&\mbox{}-\lambda_N(\lambda_N-1)\frac{f'^2(x)}{f^2(x)}
-m(m-1)\frac{h'^2(x)}{h^2(x)}-2\lambda_N\frac{f'(x)g'(x)}{f(x)g(x)}
\nonumber \\
&&\mbox{}-2m\frac{g'(x)h'(x)}{g(x)h(x)}-2m\lambda_N\frac{f'(x)h'(x)}{f(x)h(x)}
+\sum_{k=-2}^{\Vmax} V_k [f(x)]^k\bigg\}\psi_m(x),
\label{Eq6}
\eeqn
where the prime stands for the derivative with respect to $x$.
The key idea for ensuring the quasi-exact solvability of~(\ref{Eq1})
is to stabilize the space of $\psi_m(x)$ ($m=0, 1, ..., N$), or in other
words to ask for the second member of~(\ref{Eq6})
to be a linear combination of the $\psi_m(x)$.
Because of the different dependences of
the terms of this second member (with respect to $m$, etc.), we ensure the
quasi-exact solvability by requiring
\beqn
f(x)&=&\sum_{l=0}^M f_l^0[h(x)]^l,
\label{Eq7a}
\\
f'(x)&=&\sum_{l=0}^M f_l^1 [h(x)]^l,
\label{Eq7b}\\
g'(x)&=&-g(x) \sum_{l=0}^M g_l^1 [h(x)]^l,
\label{Eq8}\\
h'(x)&=&\sum_{l=0}^M h_l^1 [h(x)]^l,
\label{Eq9}
\eeqn
the different coefficients
$f_l^0$,$f_l^1$, $g_l^1$, $h_l^1$ having
to be fixed according to the potentials
we have to consider. The upper-right index indicates the derivative order
of the corresponding expanded function, e.g., to $f(x)$
correspond
%\footnote{We note that $f(x)$ must be expressed
%in term of $h(x)$, i.e., the same basis as $f'(x)$. Otherwise,
%the independency of some terms in~(\ref{Eq10}) would lead to
%a too strong
%limitation of considered potentials
%(essentially the centrifugal terms only would survive).
%For this reason, we restrict ourselves to the context}
the
expansion coefficients $f_l^0$, while to $f'(x)$ correspond the
coefficients $f_l^1$.
The upper summation index $M$ might be different for
$f(x),f'(x),g'(x),h'(x)$.
For simplicity, we take the same value: It is simply the highest
value of the three different indices, the added expansion coefficients
being
vanishing. For example, if the sum leading to $f(x)$ in Eq.~(\ref{Eq7a})
 is over  the range $l\in[0,P]$ with $P<M$, we can extend it to the
range  $l\in[0,M]$ with $f_i^0=0$ for $i\in[P+1,M]$.
Since $h(x)$ is a function to be chosen from the
beginning, $f'(x)$ and $h'(x)$ are known.
One just has to consider $M$ big enough in
order to be able to determine $g(x)$.

Taking care of~(\ref{Eq7b})--(\ref{Eq9}), we can rewrite~(\ref{Eq6}) as
\beqn
&&\left[-\frac{d^2}{dx^2} + V(x)\right] \psi_m (x) =
\Big[
-\sum_{i,j}g_j^1 g_{i-j}^1+\sum_{i,j}(i-j+1)h_j^1 g_{i-j+1}^1\nonumber \\
&&-m\sum_{i,j}jh_j^1 h_{i-j+2}^1-m(m-1)\sum_{i,j}h_j^1 h_{i-j+2}^1
+2m\sum_{i,j}
h_j^1 g_{i-j+1}^1\Big]\nonumber\\
&&\mbox{}\times g(x)[f(x)]^{\lambda_N}[h(x)]^{m+i}
+
\Big[-\lambda_N\sum_{i,j}(i-j+1)h_j^1 f_{i-j+1}^1\nonumber \\
&&\mbox{}+2\lambda_N\sum_{i,j}f_j^1 g_{i-j}^1
-2m\lambda_N\sum_{i,j}h_j^1 f_{i-j+1}^1\Big]g(x)[f(x)]^{\lambda_N-1}
[h(x)]^{m+i}\nonumber \\
&&\mbox{}+\Big[-\lambda_N (\lambda_N -1)\sum_{i,j}f_j^1 f_{i-j}^1\Big]g(x)
[f(x)]^{\lambda_N-2}[h(x)]^{m+i}\nonumber \\
&&\mbox{}+\sum_k V_k g(x)[f(x)]^{\lambda_N+k}[h(x)]^{m},
\label{Eq10}
\eeqn
where we have used
\be
\sum_{i=0}^{\infty}a_i\sum_{j=0}^{\infty}b_j=
\sum_{i=0}^{\infty}\sum_{j=0}^i b_j a_{i-j},
\label{Eq11}
\ee
 and where, for simplicity, we have omitted all the summation
indices\footnote{The relation between the coefficients $f_l^0$
and $f_l^1$ of Eqs.~(\ref{Eq7a}) and~(\ref{Eq7b}) can be
obtained with the help
of Eqs.~(\ref{Eq9}) and~(\ref{Eq11}). This relation is
$f_l^1=\sum_k h_k f_{l+1-k}^0(l+1-k)$.
}.
They will anyway not be needed in the general procedure
to be described in the next Section. For the moment,
we have not yet considered the expansion~(\ref{Eq7a}). It is used
in~(\ref{Eq10}) when extracting a common factor $f(x)^{\lambda_N-2}$:
\be
\left[-\frac{d^2}{dx^2} + V(x)\right] \psi_m (x) =
\sum_i h_{m,m+i}^{(N)} g(x)[f(x)]^{\lambda_N-2}[h(x)]^{m+i}
\label{Eq12}
\ee
with
\beqn
&&h_{m,m+i}^{(N)}\equiv -\lambda_N (\lambda_N -1)
\sum_j f_j^1 f_{i-j}^1-\lambda_N\sum_{j,l_1}(i-j-l_1+1)
 h_j^1 f_{i-j-l_1+1}^1f_{l_1}^0 \nonumber \\
&&\mbox{}+2\lambda_N\sum_{j,l_1} f_j^1 g_{i-j-l_1}^1f_{l_1}^0-2m
\lambda_N\sum_{j,l_1} h_j^1 f_{i-j-l_1+1}^1f_{l_1}^0\nonumber \\
&&\mbox{}-\sum_{j,l_1,l_2} g_j^1 g_{i-j-l_1-l_2}^1f_{l_1}^0f_{l_2}^0
+\sum_{j,l_1,l_2}(i-j-l_1-l_2+1)
h_j^1 g_{i-j-l_1-l_2+1}^1f_{l_1}^0f_{l_2}^0\nonumber \\
&&\mbox{}-m\sum_{j,l_1,l_2}j h_j^1 h_{i-j-l_1-l_2+2}^1
f_{l_1}^0f_{l_2}^0-m(m-1)\sum_{j,l_1,l_2} h_j^1
h_{i-j-l_1-l_2+2}^1f_{l_1}^0f_{l_2}^0\nonumber \\
&&\mbox{}+2m\sum_{j,l_1,l_2} h_j^1 g_{i-j-l_1-l_2+1}^1f_{l_1}^0f_{l_2}^0
\nonumber \\
&&\mbox{}+\sum_{k,l_1,l_2,...l_{k+1}} V_k f_{l_1}^0f_{l_2}^0...f_{l_{k+1}}^0
f_{i-l_1-l_2-...-l_{k+1}}^0.
\label{Eq13}
\eeqn
We thus have
\be
\left[-\frac{d^2}{dx^2} + V(x)\right] \psi_N (x) =
\sum_{m,i}c_m^{(N)} h_{m,m+i}^{(N)} g(x)[f(x)]^{\lambda_N-2}[h(x)]^{m+i}
\label{Eq14}
\ee
which, according to (\ref{Eq1}), must be equal to
\be
E_N \psi_N(x)=\sum_m c_m^{(N)} E_N \sum_{l_1,i} f_{l_1}^0f_{i-l_1}^0
g(x)[f(x)]^{\lambda_N-2}[h(x)]^{m+i}.
\label{Eq15}
\ee
Equating the respective second members of (\ref{Eq14}) and (\ref{Eq15}),
we are led to the final algebraic equations, valid for all possible
values of $j$,
giving rise to $\psi_N(x)$, as well as $E_N$:
\be
\sum_{m=0}^N c_m^{(N)} \left(h_{m,j}^{(N)}
-E_N\sum_{l_1,l_2}f_{l_1}^0f_{l_2}^0 \delta_{m,j-l_1-l_2}\right)=0,
\label{Eq16}
\ee
where, let us recall it, $h_{m,j}$ is defined through (\ref{Eq13}).
We have thus reformulated the problem~(\ref{Eq1}) to the form~(\ref{Eq16}).
The latter looks heavy to handle. However, it can be straightforwardly
implemented into a symbolic computer program.

Before turning to the general procedure,
let us finally notice that the approach
developed in this Section reduces to the one which can be found in
\cite{ref10}
if $V_{-1}=V_{-2}=0$, $\lambda_N=0$ and $f(x)=h(x)$.

\section{A more general approach}
\label{Section3}

In this Section, we briefly describe the most general
algorithm we have developped starting from the approach of the previous
Section. As far as we now, there is no extension of it if we
restrict ourselves to one basis function $h(x)$. A generalization
to more basis functions ($h_1(x),h_2(x), ...$) is possible.
In particular, some of the potentials we have studied may be treated using
this generalization. However, they may as well be treated using
the algorithm we present below. For this reason,  the study of the
generalized algorithm based on more basis functions is left
for a future work.

In the previous Section, we have described in details the philosophy
of our approach to find analytical solutions to a given \Sch equation.
Those analytical solutions are possible if we can factor out
appropriate powers of $h(x)$ and $f(x)$, thus
closing the system, see Eq.~(\ref{Eq12}).
This later equation was found using the expansion of the potential on
a basis $f(x)$,
see Eq.~(\ref{Eq2}), together with the assumptions~(\ref{Eq7a})--(\ref{Eq9}).
We show now how to relax these constraints in order to
treat the maximum number of different potentials.

The simplest extension is to allow negative indices for
the sums in Eqs.~(\ref{Eq7a})--(\ref{Eq9}). This
brings no difficulties, and this is why we did not specified the range
of the indices in the previous Section:
\be
\sum_{i=-\infty}^{\infty}a_i\sum_{j=-\infty}^{\infty}b_j=
\sum_{i=-\infty}^{\infty}\sum_{j=-\infty}^{\infty} b_j a_{i-j}
\equiv \sum_i\sum_j b_j a_{i-j},
\label{Eq17}
\ee
where, from now on, an index without range means a running on ${\mathbb Z}$.
Taking care of negative indices is required for example when using a
basis such as $h(x)=\exp(-x)$. One can understand that  to obtain
a solution starting like
$\cosh(x)$ may require negative powers of $h(x)$:
$2\cosh(x)=h(x)+h(x)^{-1}$.
For the same
reason, the expansion of~(\ref{Eq3}) is extended to start from
negative orders.

To allow more flexibility in the choice of the
potential $V(x)$ in~(\ref{Eq2}),
it is also necessary to make the sum begining to a lower negative index
than $k=-2$. For this reason, we make it begin at $k=-\Vmin$, with the
condition $\Vmin\ge 2$ this restriction being due to the fact that
at least a factor of $f(x)^{\lambda_N-2}$ must be factored out in order to
close the system.

In studying the way the different powers of $h(x)$ enter into
Eq.~(\ref{Eq6}), we can also convince ourselves that
 integer powers of the square root of the right-hand-side
of Eq.~(\ref{Eq9}) can be chosen. The analysis can be done
using an arbitray power $\alpha$:
$h'(x)=\left\{\sum_{l=-\infty}^{\infty} h_l^1 [h(x)]^l\right\}^{\alpha}$.
Plugging this expansion into~(\ref{Eq6}), it is straighforward to see
that the condition $2\alpha=n\in{\mathbb N}$ is required. Moreover,
using~(\ref{Eq17}) and properly redefining the expansion coefficients,
we can always write this $h'(x)$ as the product of a series
of the form of~(\ref{Eq9})
with a series of the form of the square root of~(\ref{Eq9}).

Using a square root in $h'(x)$ also requires to generalize the term
$g(x)$ in order, for example, to treat properly the term $g'(x)h'(x)$
in Eq.~(\ref{Eq6}). The generalization consists in using the exponential
of a $h(x)$-series, instead of the differential form given in~(\ref{Eq8}),
or, equivalently, to multiply the latter by $h'(x)$ and properly
redefine the expansion coefficients.

To summarize, the different expansions replacing Eqs.~(\ref{Eq2}),
(\ref{Eq7a})
and~(\ref{Eq8})--(\ref{Eq9}) are:
\beqn
\psi_N(x)&=&\sum_{m=-N}^Nc_m^{(N)}\psi_m(x),
\label{Eq17b}
\\
V(x)&=&\sum_{k=-\Vmin}^{\Vmax}V_k[f(x)]^k,
\label{Eq18}\\
f(x)&=&\sum_l f_l^0[h(x)]^l,
\label{Eq19}
\\
g(x)&=&\exp\left\{-\sum_l \tg_l^1 [h(x)]^l-\tg_{\log}^1\ln[h(x)]\right\},
\label{Eq20}\\
h'(x)&=&\left\{\sum_l \sth_l^1 [h(x)]^l\right\}
\left\{\sum_l \dth_l^1 [h(x)]^l\right\}^{1/2},
\label{Eq21}
\eeqn
where we have added a term $\tg_{\log}^1\ln[h(x)]$ in Eq.~(\ref{Eq20}).
This is equivalent to multiplying Eq.~(\ref{Eq3}) by  $h(x)^{-\tg_{\log}^1}$,
thus generalizing again the approach of the previous Section, and
increasing its flexibility. We recover the algorithm of the previous
Section by taking $c_{m<0}^{(N)}=0,
% \sth_l^1=h_l^1,
\dth_l^1=\delta_{l,0},$
 $\tg_{\log}^1=0$, $g_l^1=\sum_nh_n^1\tg_{l-n+1}^1(l-n+1)$.
We also remark that
not all the unknowns are independent: For example, a problem with $N=1,
\tg_{\log}^1=0$ implies the
sum, see Eq.~(\ref{Eq17b}),
$c_{-1}^{(1)}h(x)^{-1}+c_0^{(1)}+c_1^{(1)}h(x)$.
 This is equivalent to $h(x)^{-1}[c_{-1}^{(1)}+c_0^{(1)}h(x)+c_1^{(1)}h(x)^2]$.
A similar result is achieved with $N=2,\tg_{\log}^1=1,c_{-2}^{(2)}
=c_{-1}^{(2)}=0,c_{0}^{(2)}=c_{-1}^{(1)},c_{1}^{(2)}=c_{0}^{(1)}
,c_{2}^{(2)}=c_{1}^{(1)}$.

The last possible generalization we quote is to modify again the expansion
of the potential. The function $f(x)$ in Eq.~(\ref{Eq18}) can be modulated
by any linear combination of the basis function $h(x)$. Indeed, with
new coefficients $\vf_{l,k}$ of two indices
refering to the power $k$ of $f(x)$
and to the power $l$ of $h(x)$, the potential
\be
V(x)=
\sum_{k=-\Vmin}^{\Vmax}V_k[f(x)]^k\sum_l \vf_{l,k}[h(x)]^l,
\label{Eq22}
\ee
when multiplied by $\psi_m(x)$ of Eq.~(\ref{Eq5}) just leads to the
factorization
of a power of $h(x)$. This is seen as follows:
\beqn
V(x)\psi_m(x)&=&\nonumber\\
&&\hspace{-3cm}\sum_{k=0}^{\Vmin+\Vmax}
V_{k-\Vmin}f(x)^k\sum_l \vf_{l,k-\Vmin}[h(x)]^l
g(x)[h(x)]^mf(x)^{\lambda_N-\Vmin}\nonumber\\
&&\hspace{-3cm}=\sum_{k=0}^{\Vmin+\Vmax}V_{k-\Vmin}\sum_n {}^k\!f_n^0[h(x)]^n
\sum_l \vf_{l,k-\Vmin}[h(x)]^l
g(x)[h(x)]^mf(x)^{\lambda_N-\Vmin}\nonumber\\
&&\hspace{-3cm}=\sum_{k=0}^{\Vmin+\Vmax}V_{k-\Vmin}\sum_l {}^k\!\tilde{f}_l^0
[h(x)]^{l+m}g(x)f(x)^{\lambda_N-\Vmin},
\label{Eq23}
\eeqn
where we have used---compare with the last line of Eq.~(\ref{Eq13})---
\be
f(x)^k=\left\{\sum_l f_l^0[h(x)]^l\right\}^k=\sum_l {}^k\!f_l^0[h(x)]^l,
\label{Eq24}
\ee
and where
\goodbreak
\beqn
\sum_n {}^k\!f_n^0[h(x)]^n
\sum_l \vf_{l,k-\Vmin}[h(x)]^l&=&\sum_l[h(x)]^l\sum_n{}^k\!f_n^0
\vf_{l-n,k-\Vmin}\nonumber\\
&=& \sum_l{}^k\!\tilde{f}_l^0[h(x)]^l,
\label{Eq25}
\eeqn
the last equality defining the coefficient ${}^k\!\tilde{f}_l^0$.
\goodbreak
In Eq.~(\ref{Eq24}), the coefficients ${}^k\!f_l^0$ are defined recursivelly:
\be
f(x)^k=f(x)f(x)^{k-1}\Longrightarrow {}^k\!f_l^0=
\sum_n {}^{k-1}\!f_{l-n}^0 f_n^0,
\label{Eq26}
\ee
with the starting coefficient ${}^0\!f_l^0=\delta_{l,0}$. In defining the
different coefficients, we have taken the convention that
the upper-left index refers to the power of the given function, while, as in
the previous Section, the upper-right index indicates the order
of its derivative. (We do not, however,
specify the upper-left index when it is equal to 1, e.g., ${}^1\!f_l^i\equiv
f_l^i$.)

%As an example of the generalized form of the potential~(\ref{Eq22}),
%we can mention the screened Coulomb potential \cite{znojil84} for which
%$V(x)=L(L+1)/r^2-\alpha/(r+c)$, the first term being the centrifugal
%barrier. It enters our
%formalism if we take $f(x)=(x+c)$ and $h(x)=x$
%with $V_{i}=\vf_{l,k}=0$ except for $V_{-1}=-\alpha,V_0=L(L+1)$ together with
%$\vf_{l,-1}=\delta_{l,0}$ and $\vf_{l,0}=\delta_{l,-2}$.

With the generalization given in Eqs.~(\ref{Eq19})--(\ref{Eq22}), and
with the definitions of the coefficients in~(\ref{Eq25})--(\ref{Eq26}),
we are now prepared to give the full algorithm solving Eq.~(\ref{Eq1}).
For this, we introduce also the following notation for the different terms
entering Eq.~(\ref{Eq6}):
\beqn
{}^2\!\sth_l^1&\equiv&\sum_m \sth_m^1\dth_{l-m}^1,\label{Eq27}\\
\left.[h'(x)]^2\right.&=&\sum_l[h(x)]^l
\sum_m{}^2\!\sth_m^1\dth_{l-m}^1\equiv
\sum_l{}^2\!h_l^1[h(x)]^l,\\
f'(x)&=&\sum_l f_{l+1}^0(l+1)[h(x)]^lh'(x)\equiv \sum_l\tf_l^1[h(x)]^lh'(x),\\
{}^2\!\tf_l^1&\equiv&\sum_m\tf_m^1\tf_{l-m}^1,\\
\left.[f'(x)]^2\right.&=&\sum_l[h(x)]^l\sum_m{}^2\!h_m^1{}^2\!\tf_{l-m}^1
\equiv\sum_l {}^2\!f_l^1[h(x)]^l,\\
g'(x)&=&-g(x)\left\{\sum_l\tg_{l+1}^1(l+1)[h(x)]^l+\tg_{\log}^1/h(x)
\right\}h'(x)\nonumber\\
&\equiv&
-g(x)h'(x)\sum_lg_l^1[h(x)]^l,\label{Eq28}\\
{}^2\!\tg_l^1&\equiv&\sum_mg_m^1g_{l-m}^1,\\
\left.[g'(x)]^2\right.&=&[g(x)]^2\sum_l[h(x)]^l\sum_m{}^2\!h_m^1
{}^2\!\tg_{l-m}^1\equiv g(x)^2\sum_l{}^2\!g_l^1[h(x)]^l,\\
f'(x)h'(x)&=&\sum_l[h(x)]^l\sum_m{}^2\!h_m^1\tf_{l-m}^1\equiv
\sum_l\fh_l^1[h(x)]^l,\\
g'(x)h'(x)&=&-g(x)\sum_l[h(x)]^l\sum_m{}^2\!h_m^1g_{l-m}^1\equiv
\sum_l\gh_l^1[h(x)]^l,\\
f'(x)g'(x)&=&-g(x)[h'(x)]^2\sum_l[h(x)]^l\sum_mg_m^1\tf_{l-m}^1\equiv
-g(x)[h'(x)]^2\sum_l\tilde{\fg}_l^1[h(x)]^l\nonumber\\
&=&-g(x)\sum_l[h(x)]^l\sum_m\tilde{\fg}_m^1{}^2\!h_{l-m}^1
\equiv
-g(x)\sum_l\fg_l^1[h(x)]^l,\\
\tth_l^1&\equiv&\sum_m\sth_{m+1}^1\sth_{l-m}^1,\\
h''(x)&=&\sum_l[h(x)]^l\sum_m
\left[\tth_m^1\dth_{l-m}^1+\frac{1}{2}\dth_{m+1}(m+1){}^2\!\sth_{l-m}^1\right]
\nonumber\\
&\equiv&\sum_lh_l^2[h(x)]^l,\\
f''(x)&=&\sum_l[h(x)]^l\sum_m\left[{}^2\!h_m^1\tf_{l+1-m}^1(l+1-m)
+h_m^2\tf_{l-m}^1\right]\nonumber\\
&\equiv&\sum_lf_l^2[h(x)]^l,\\
g''(x)&=&g(x)\sum_l[h(x)]^l\left[
{}^2\!g_l^1-\sum_mg_{m+1}^1(m+1){}^2\!h_{l-m}^1-\sum_mg_m^1h_{l-m}^2
\right]\nonumber\\
&\equiv&g(x)\sum_lg_l^2[h(x)]^l,\label{Eq29}
\eeqn
where the equivalence signs in the above equations define new coefficients.
In obtaining these, we have made repetive use of~(\ref{Eq17}). Note that
the coefficient $g_l^1$ of Eq.~(\ref{Eq28})
is defined differently from~(\ref{Eq8}).
These notations are quite lengthy. However, they allow to obtain
an algebraic equation very suitable for
a symbolic computer program, generalization
of Eq.~(\ref{Eq16}) with~(\ref{Eq13}):
\beqn
&&\hspace{-1.5cm}\mbox{}\sum_{m=-N}^Nc_m^{(N)}
\left[h_{m,j}^{(N)}-E_N\left({}^{\Vmin}\!f_{j-m}^0\right)\right]=0
\label{Eq31},\\
&&\hspace{-1.5cm}\mbox{}h_{m,j}^{(N)}\equiv
-\sum_n
\Big[\lambda_N(\lambda_N-1)\left({}^{\Vmin-2}\!f_n^0\right)
\left({}^2\!f_{j-m-n}^1\right)
+\lambda_N\left({}^{\Vmin-1}\!f_n^0\right)
f_{j-m-n}^2\nonumber\\
&&\hspace{.5cm}\mbox{}-2\lambda_N\left({}^{\Vmin-1}\!f_n^0\right)
\fg_{j-m-n}^1
+2m\lambda_N\left({}^{\Vmin-1}\!f_n^0\right)
\fh_{j-m-n+1}^1\nonumber\\
&&\hspace{.5cm}\mbox{}+\left({}^{\Vmin}\!f_n^0\right)
g_{j-m-n}^2
+m\left({}^{\Vmin}\!f_n^0\right)h_{j-m-n+1}^2\nonumber\\
&&\hspace{.5cm}\mbox{}+m(m-1)\left({}^{\Vmin}\!f_n^0\right)
\left({}^2\!h_{j-m-n+2}^1\right)
-2m\left({}^{\Vmin}\!f_n^0\right)\gh_{j-m-n+1}^1\Big]\nonumber\\
&&\hspace{.5cm}\mbox{}+\sum_{k=0}^{\Vmin+\Vmax}
V_{k-\Vmin}\left({}^k\!\tilde{f}_{j-m}^0\right).
\label{Eq30}
\eeqn
This algebraic equation has to be solved for each value of $j$.
Apart from the possible generalization using several basis
functions $(h_1(x),h_2(x),\cdots)$ that we mentioned in the beginning
of this Section,
this equation
is the most general algorithm we have found\footnote{A slight
generalization is possible, consisting in taking, instead of the 
parametrization~(\ref{Eq22}) of the potential, the ratio
of two such paramatrizations. We do not consider this generalization here
because we did not find an example needing it. This does not, however,
exclude this possibility.} 
to solve Eq.~(\ref{Eq1}).

In the next Section, we present several examples using both the generalized
algorithm~(\ref{Eq30})--~(\ref{Eq31}) and the simplest one given
by Eqs.~(\ref{Eq13}) and~(\ref{Eq16}). We use the latter when
that the generalizations are not needed because it works faster.
 The examples are treated using a Mathematica
implementation of Eqs.~(\ref{Eq13}), (\ref{Eq16}),
(\ref{Eq30}) and~(\ref{Eq31}).

\section{Examples}
\label{Section4}

In this Section, we prove the convenience of our method for the determination
of the  eigenvalues and eigenfunctions of well-known Q.E.S. equations.
We start with the genelarized sextic oscillator. We begin with a general
set of parameters and show by fixing some of these parameters that
the constraints between the different left parameters can be solved.
In order to illustrate our method, a lot of details are given in the
first treatment of the problem. Then, as we go on, only the main
points are stressed.

\subsection{The generalized sextic oscillator}

The generalized sextic potential has the form
\beqn
&&V(x)=V_3b^3x^6+(3V_3ab^2+V_2b^2)x^4+(3V_3a^2b+2V_2ab+V_1b)x^2 \nonumber \\
&&\hspace{1cm}\mbox{}+V_3a^3+V_2a^2
+V_1a+V_0+\frac{V_{-1}}{a+bx^2}+\frac{V_{-2}}{(a+bx^2)^2}
\label{Eq46}
\eeqn
with $bV_3\ge0$ and $ab\ge0$.
It coincides with the sextic radial oscillator
analyzed by Ushveridze \cite{ref4} if
\beqn
&&V_3=t^2, V_2=0,V_1=-4t\left(s+\frac{1}{2}+n
\right),V_0=0, \nonumber \\
&&V_{-1}=4\left(s-\frac{1}{4}\right)
\left(s-\frac{3}{4}\right),V_{-2}=0,a=0,b=1,
\label{Eq47}
\eeqn
while the Tkachuk potential \cite{ref5} is recovered for
\be
V_3=\frac{1}{4b},V_2=-\frac{a}{4b},V_1=-3,V_0=\frac{5}{2}a,
V_{-1}=\frac{3}{4}b,V_{-2}=\frac{3}{4}ab
\label{Eq48}
\ee
in the context of two solutions.

We see from~(\ref{Eq46}) that $V(x)$
can be written as Eq.~(\ref{Eq2}) with the identification
\be
M=3, f(x)=a+bx^2.
\ee
This, as well as $f'(x)=2bx$, clearly suggests the basis
function $h(x)\equiv x$.
From the relations~(\ref{Eq7a})--(\ref{Eq9}) we get the nonvanishing
coefficients:
\be
f_0^0=a,f_2^0=b, f_1^1=2b, h_0^1=1.
\ee
The unknowns of the problem are $E_N, \lambda_N, c_m^{(N)}, g_l^1$. They
have to be found from Eq.~(\ref{Eq16}). Studying the case $N=0$, we can
convince ourselves that the only nonvanishing coefficients of the weight
function $g(x)$ are
 $g_l^1 \; (l=0,1,2,3)$.
Indeed, Eq.~(\ref{Eq16}) reduces to
\be
h_{0,j}^{(0)}-E_0\left(a^2\delta_{j,0}+2ab\delta_{j,2}
+b^2\delta_{j,4}\right)=0,\; j=0,1,...,10
\label{Eq51}
\ee
with---see Eq.~(\ref{Eq13})---
\beqn
h_{0,j}^{(0)}&=&\left[-2ab\lambda_0-a^2(g_0^1)^2+a^2g_1^1
+V_{-2}+aV_{-1}+a^2V_0+a^3V_1+a^4V_2
+a^5V_3\right]\delta_{j,0} \nonumber \\
&&\mbox{}+\left(4ab\lambda_0g_0^1-2a^2g_0^1g_1^1+2a^2g_2^1\right)
\delta_{j,1}+\left[-4b^2\lambda_0^2+2b^2\lambda_0\right. \nonumber \\
&&\mbox{}+4ab\lambda_0g_1^1-2a^2g_0^1g_2^1-a^2(g_1^1)^2-2ab(g_0^1)^2
+3a^2g_3^1+2abg_1^1
+V_{-1}b+2abV_0 \nonumber \\
&&\mbox{}\left.+3a^2bV_1+4a^3bV_2+5a^4bV_3\right]\delta_{j,2}
+\left(4ab\lambda_0g_2^1+4b^2\lambda_0g_0^1\right. \nonumber \\
&&\mbox{}\left.-2a^2g_0^1g_3^1-2a^2g_1^1g_2^1-4abg_0^1g_1^1
+4abg_2^1\right)\delta_{j,3}
+\left[4ab\lambda_0g_3^1+4b^2\lambda_0g_1^1\right. \nonumber \\
&&\mbox{}-2a^2g_1^1g_3^1-a^2(g_2^1)^2-4abg_0^1g_2^1-2ab(g_1^1)^2
-b^2(g_0^1)^2
+6abg_3^1+b^2g_1^1+b^2V_0 \nonumber \\
&&\mbox{}\left.+3ab^2V_1+6a^2b^2V_2+10a^3b^2V_3\right]\delta_{j,4}
+\left(4b^2\lambda_0g_2^1-2a^2g_2^1g_3^1\right. \nonumber \\
&&\mbox{}\left.-4abg_0^1g_3^1-4abg_1^1g_2^1
-2b^2g_0^1g_1^1+2b^2g_2^1\right)\delta_{j,5}
+\left[4b^2\lambda_0g_3^1-a^2(g_3^1)^2\right. \nonumber \\
&&\mbox{}-4abg_1^1g_3^1-2ab(g_2^1)^2-2b^2g_0^1g_2^1-b^2(g_1^1)^2
+3b^2g_3^1+b^3V_1+4ab^3V_2 \nonumber \\
&&\mbox{}\left.+10a^2b^3V_3\right]\delta_{j,6}+
\left(-4abg_2^1g_3^1-2b^2g_0^1g_3^1-2b^2g_1^1g_2^1\right)
\delta_{j,7}\nonumber \\
&&\mbox{}+\left[-2ab(g_3^1)^2-2b^2g_1^1g_3^1-b^2(g_2^1)^2
+b^4V_2+5ab^4V_3\right]\delta_{j,8}\nonumber \\
&&\mbox{}+\left(-2b^2g_2^1g_3^1\right)\delta_{j,9}
+\left[-b^2(g_3^1)^2+b^5V_3\right]\delta_{j,10}.
\label{Eq52}
\eeqn
With vanishing $g_l^1, l=5,...$ expansion coefficients, we would have get
an extra term $g_4^1\delta_{j,11}$, implying $g_4^1=0$. Repeating
this argument recursively, we see that only $g_l^1, l=0,1,2,3$ survive.

From the values $j=10,9,8,7$ we  compute the weight function:
\beqn
g(x)&\equiv&\exp\left(-\sum_{l=1}^4 \frac{g_{l-1}^1}{l}x^l\right)
 \nonumber \\
&=&\exp\left(-\frac{3}{4}a\sqrt{bV_3}x^2-\frac{1}{4}\sqrt{\frac{b}{V_3}}V_2x^2
-\frac{1}{4}\sqrt{b^3V_3}x^4\right),
\label{Eq53}
\eeqn
while for $j=10$ we have taken the positive root $g_3^1=\sqrt{b^3V_3}$.
We could have chosen the negative root. This would have lead
to a nonphysical solution because it is not bounded when $x\rightarrow\infty$.
We note however that unbounded solutions have their usefulness: they can
be used to generate new potentials and the corresponding (bounded)
eigenfunctions
through Darboux transformations \cite{dsv}.

The index $j=6$ fixes
$\lambda_0$. Together, the wavefunction $\psi_0(x)$
is determined  to be
\beqn
\psi_0(x)&=&c_0\left(a+bx^2\right)^{-\frac{3}{16}
\frac{a^2}{\sqrt{b}}\sqrt{V_3}
-\frac{1}{8}\frac{a}{\sqrt{b}}
\frac{V_2}{\sqrt{V_3}}+\frac{1}{16}\frac{1}{\sqrt{b}}
\frac{V_2^2}{V_3^{\frac{3}{2}}}-\frac{3}{4}-\frac{1}{4}\frac{V_1}{\sqrt{bV_3}}
} \nonumber\\
&&\hspace{0.5cm}\mbox{}\times
\exp\left(-\frac{3}{4}a\sqrt{bV_3}x^2-\frac{1}{4}
\sqrt{\frac{b}{V_3}}V_2x^2
-\frac{1}{4}\sqrt{b^3V_3}x^4\right),
\label{Eq54}
\eeqn
the quantity $c_0$ being fixed by normalization.
The next significant value $j=4$ leads to the knowledge of $E_0$
\be
E_0=\frac{5}{8}a^3V_3+\frac{3}{8}a^2V_2+\frac{1}{2}aV_1
+V_0-\frac{1}{8}a\frac{V_2^2}{V_3}+\frac{1}{8}\frac{V_2^3}{V_3^2}
-\frac{1}{2}\frac{V_1V_2}{V_3}-\sqrt{\frac{b}{V_3}}V_2.
\label{Eq55}
\ee
Finally, taking care of $j=2$ and $j=0$, we are left with
two constraints on the potential~(\ref{Eq46}):
\beqn
V_{-1}&=&\frac{1}{64}\frac{V_2^4}{V_3^3}-\frac{1}{8}\frac{V_1V_2^2}{V_3^2}
+\frac{1}{4}\frac{V_1^2}{V_3}-\frac{1}{2}\sqrt{b}
\frac{V_2^2}{V_3^{\frac{3}{2}}}+2\sqrt{b}\frac{V_1}{\sqrt{V_3}}
+\frac{15}{4}b-\frac{1}{2}a\sqrt{b}\frac{V_2}{\sqrt{V_3}} \nonumber \\
&&\hspace{-1.5cm}\mbox{}
+\frac{1}{16}a\frac{V_2^3}{V_3^2}-\frac{1}{4}a\frac{V_1V_2}{V_3}
-\frac{1}{8}a^2V_1-\frac{7}{16}a^3V_2
-\frac{5}{32}a^2\frac{V_2^2}{V_3}-\frac{15}{64}a^4V_3,
\label{Eq56}\\
V_{-2}&=&\frac{1}{8}a\frac{V_1V_2^2}{V_3^2}
-\frac{1}{4}a\frac{V_1^2}{V_3}-\frac{1}{64}a\frac{V_2^4}{V_3^3}
-\frac{1}{4}a^2\frac{V_1V_2}{V_3}+\frac{1}{16}a^2\frac{V_2^3}{V_3^2}
-\frac{9}{64}a^5V_3 \nonumber \\
&&\hspace{-1.5cm}\mbox{}-\frac{3}{16}a^4V_2
-\frac{3}{8}a^3V_1-\frac{5}{2}a\sqrt{b}\frac{V_1}{\sqrt{V_3}}
-\frac{21}{4}ab+\frac{1}{32}a^3\frac{V_2^2}{V_3}
+\frac{5}{8}a\sqrt{b}\frac{V_2^2}{V_3^{\frac{3}{2}}} \nonumber \\
&&\hspace{-1.5cm}\mbox{}-\frac{5}{4}a^2\sqrt{b}\frac{V_2}{\sqrt{V_3}}
-\frac{15}{8}a^3\sqrt{b}\sqrt{V_3},
\label{Eq57}
\eeqn
respectively. The different choices~(\ref{Eq47}) (with $n=0$)
and~(\ref{Eq48}) are
compatible with these contraints, which are
 limitations on the potential: There is at least one solution
to the \Sch equation~(\ref{Eq1}) with~(\ref{Eq46}),
this solution being given by
Eq.~(\ref{Eq54}), only if the different $V_i,i=-2,...,3$ coefficients of the
potential satisfy the constraints.

We proceed in a similar way for the $N=1$-case.
Now, the set of  equations~(\ref{Eq16}) reads, for
$j=0,...,11$:
\beqn
&&c_0^{(1)}\left[h_{0,j}^{(1)}-E_1\left(a^2\delta_{j,0}
+2ab\delta_{j,2}+b^2\delta_{j,4}\right)\right]\nonumber \\
&&\hspace{1cm}\mbox{}+c_1^{(1)}\left[h_{1,j}^{(1)}-E_1
\left(a^2\delta_{j,1}+2ab\delta_{j,3}+b^2\delta_{j,5}\right)\right]
=0.
\label{Eq58}
\eeqn
The quantity $h_{0,j}^{(1)}$ is again given by
Eq.~(\ref{Eq52}) up to the replacement of
$\lambda_0$ by $\lambda_1$, while $h_{1,j}^{(1)}$ is
\beqn
h_{1,j}^{(1)}&=&2a^2g_0\delta_{j,0}
+\left[-6ab\lambda_1-a^2(g_0^1)^2+3a^2g_1^1+V_{-2}+aV_{-1}
+a^2V_0+a^3V_1\right.\nonumber \\
&&\hspace{-1.5cm}\mbox{}\left.+a^4V_2+a^5V_3\right]\delta_{j,1}
+\left(4ab\lambda_1g_0^1-2a^2g_0^1g_1^1+4a^2g_2^1+4abg_0^1\right)\delta_{j,2}
 \nonumber \\
&&\hspace{-1.5cm}\mbox{}+\left[-4b^2\lambda_1^2-2b^2\lambda_1
+4ab\lambda_1g_1^1-2a^2g_0^1g_2^1-a^2(g_1^1)^2-2ab(g_0^1)^2+5a^2g_3^1
+6abg_1^1\right.\nonumber \\
&&\hspace{-1.5cm}\mbox{}\left.+V_{-1}b+2abV_0
+3a^2bV_1+4a^3bV_2+5a^4bV_3\right]
\delta_{j,3}+\left(4ab\lambda_1g_2^1+4b^2\lambda_1g_0^1\right. \nonumber \\
&&\hspace{-1.5cm}\mbox{}\left.-2a^2g_0^1g_3^1-2a^2g_1^1g_2^1-4abg_0^1g_1^1
+8abg_2^1+2b^2g_0^1\right)\delta_{j,4}+\left[4ab\lambda_1g_3^1
+4b^2\lambda_1g_1^1\right. \nonumber \\
&&\hspace{-1.5cm}\mbox{}-2a^2g_1^1g_3^1-a^2(g_2^1)^2-4abg_0^1g_2^1
-2ab(g_1^1)^2-b^2(g_0^1)^2+10abg_3^1+3b^2g_1^1+b^2V_0 \nonumber \\
&&\hspace{-1.5cm}\mbox{}\left.+3ab^2V_1+6a^2b^2V_2
+10a^3b^2V_3\right]\delta_{j,5}+\left(4b^2\lambda_1g_2^1
-2a^2g_2^1g_3^1\right. \nonumber \\
&&\hspace{-1.5cm}\mbox{}\left.-4abg_0^1g_3^1-4abg_1^1g_2^1-2b^2g_0^1g_1^1
+4b^2g_2^1\right)\delta_{j,6}
+\left[4b^2\lambda_1g_3^1-a^2(g_3^1)^2\right. \nonumber \\
&&\hspace{-1.5cm}\mbox{}-4abg_1^1g_3^1-2ab(g_2^1)^2-2b^2g_0^1g_2^1
-b^2(g_1^1)^2
+5b^2g_3^1+b^3V_1+4ab^3V_2 \nonumber \\
&&\hspace{-1.5cm}\mbox{}\left.+10a^2b^3V_3
\right]\delta_{j,7}+\left(-4abg_2^1g_3^1-2b^2g_0^1g_3^1
-2b^2g_1^1g_2^1\right)\delta_{j,8}\nonumber \\
&&\hspace{-1.5cm}\mbox{}+\left[-2ab(g_3^1)^2-2b^2g_1^1g_3^1
-b^2(g_2^1)^2+b^4V_2+5ab^4V_3\right]\delta_{j,9}\nonumber \\
&&\hspace{-1.5cm}\mbox{}+\left(-2b^2g_2^1g_3^1\right)\delta_{j,10}
+\left[-b^2(g_3^1)^2+b^5V_3\right]\delta_{j,11}.
\label{Eq59}
\eeqn
The analysis of the different values of
$j$ leads to the following eigenfunction
\be
\psi_1(x)=c_1 \frac{x}{\left(a+bx^2\right)^{\frac{1}{2}}} \psi_0(x)
\label{Eq60}
\ee
corresponding to the energy
\be
E_1=E_0+2a\sqrt{bV_3}
\label{Eq61}
\ee
iff the constraints on $V(x)$
\beqn
V_{-1}(N=1)&=&
V_{-1}(N=0)-a\sqrt{b}\frac{V_2}{\sqrt{V_3}}-a^2\sqrt{bV_3},
\label{Eq62}\\
V_{-2}(N=1)&=&V_{-2}(N=0)-6ab-a\sqrt{b}\frac{V_1}{\sqrt{V_3}}
+\frac{1}{4}a\sqrt{b}\frac{V_2^2}{V_3^{\frac{3}{2}}}
\nonumber\\
&&\hspace{3cm}\mbox{}-\frac{1}{2}a^2\sqrt{b}\frac{V_2}{\sqrt{V_3}}
-\frac{3}{4}a^3\sqrt{bV_3}
\label{Eq63}
\eeqn
are satisfied.

We have also computed the cases $N=2,3,4$. The results are too cumbersome
to be written here. We have however observed that the constraints
always change with $N$ and that to $\psi_N(x)$ corresponds the $N$th
excited state (the identification is done looking at the number
of nodes of the wavefunction).
Since the constraints change with $N$, this  means that
each  obtained level corresponds to a different potential.
This is at variance with the particular case  $a=0$
(see Ushveridze \cite{ref4}). In this limit, we see from the comparison
between Eqs.~(\ref{Eq54})--(\ref{Eq57}) and Eqs.~(\ref{Eq60})--(\ref{Eq63})
that the cases $N=0$ and $N=1$ coincide.

We also see
quite straightforwardly that the constraints~(\ref{Eq56})--(\ref{Eq57})
are compatible with the parametrization given in Eq.~(\ref{Eq47}), provided
 $n=0$. In fact, going to higher integer $n$, we can
convince ourselves that there are $n+1$ solutions compatible with the
constraint on $V_{-1}$. These solutions are defined on the half line
$]0,\infty[$, due to the centrifugal
barrier $V_{-1}/x^2$, hence the name sextic {\em radial} oscillator.
The higher $n$ solutions are investigated using higher $N$ values in our
algorithm.

In general, playing with the different parameters, we can generate a lot
of different solutions:
\begin{itemize}
\item In the general case above, we mentioned the
fact that to one potential corresponds a single solution;
\item To the
sextic radial oscillator corresponds a given number of solutions, depending
on the value of the integer $n$ in~(\ref{Eq47});
\item Eliminating  also
the centrifugal barrier of the sextic radial oscillator, $V_{-1}=0$, we
are left with a model defined on $]-\infty,\infty[$, hence the
name sextic oscillator. In that case, the exact solutions fall
into parity classes. To see this statement, the parametrization~(\ref{Eq47})
is replaced by
\beqn
&&V_3=t^2, V_2=0,V_1=-4t\left(n+\frac{p}{2}+\frac{3}{4}
\right),V_0=0, \nonumber \\
&&V_{-1}=0,V_{-2}=0,a=0,b=1,
\label{Eq64}
\eeqn
where $p$ denotes the parity. For a given $n$, one may get series
of solutions: One with $p=0$ (even solutions), and one with $p=1$ (odd
solutions). Since $p$ can take two values, the potential is then different
for odd and even solutions: For a given potential, the solutions
fall into parity classes, as said before;
\item
Even with $V_{-1}$ and $V_{-2}$ nonvanishing, we can have
an exactly solvable potential. For example, the potential
\be
V(x)=\frac{x^2}{4}+\frac{4}{1+x^2} -\frac{8}{(1+x^2)^2}
\label{Eq65}
\ee
is exactly solvable. This can be shown easily with our approach because
we get no constraint. Theoretically, this is even more easily
seen by noting that this potential can be
constructed from the harmonic oscillator using a Darboux transformation
\cite{dsv};
\item The Kuliy-Tkachuk potential \cite{ref7} obtained from the general case
with the special parametrization~(\ref{Eq48}) is also
interesting in the sense that it always generates  a given number of solutions.
This number is fixed from the start (for the Kuliy-Tkachuk potential, this
number is three).
\end{itemize}

In the following, we shall review the different possibilities we have just
enumerated.

\subsubsection{The general case}

We have already discussed the general case given by
Eq.~(\ref{Eq46}) for $N=0$ and $N=1$. For higher $N$, we get,
before using the parametrization~(\ref{Eq47})
which still has to be shown to be consistent, the
condition $\lambda_N=\lambda_0-N/2$. The equations for the constraints
relating $V_{-1}$ and $V_{-2}$ to the other coefficients,
and the equations for the coefficients $c_m^{(N)}$, relating them
to a single one to be fixed by normalization, for instance $c_0^{(N)}$, are
too complicated to be solved analytically for arbitrary $N$.
However, we can find at least some
of the solutions and see what they imply.
Using our algorithm, one can see by direct substitution in the corresponding
constraint equation that $V_{-1}^{(2N)}=V_{-1}^{(0)};
V_{-2}^{(2N)}=V_{-2}^{(0)}$
and $V_{-1}^{(2N+1)}=V_{-1}^{(1)}; V_{-2}^{(2N+1)}=V_{-2}^{(1)}$ are
compatible constraints, for which we also observe that $E_{2N}=E_0$ and
$E_{2N+1}=E_1$. This however does not lead to new results: In fact, we just
recover the case $N=0$ and $N=1$, respectively. This is seen as follows,
taking the case $N=2$ as an example:
Using the constraints~(\ref{Eq56}) and~(\ref{Eq57}), we can
solve the equation for $c_2^{(2)}$ as a function of $c_0^{(2)}$. We get
$c_2^{(2)}=c_0^{(2)}b/a$. Together with the relations
$\lambda_N=\lambda_0-1$ and
$f(x)\equiv (a+bx^2)$, we have $f(x)^{\lambda_2}(c_0^{(2)}+c_0^{(2)}x^2)
=f(x)^{\lambda_0}c_0^{(2)}/a$. Then, the wavefunction $\psi_2$ and $\psi_0$
are equal, up to an unimportant constant.
This analysis remains true to higher $N$. This means that, when taking
higher $N$, we cannot take the same constraint equations for $V_{-1}$ and
$V_{-2}$ as the equations for $N=0$ or $N=1$ to get new results. This implies
that the constraints on $V_{-1}$ and
$V_{-2}$ are different for each value of $N$, a result that we mentioned
earlier.
%The fact that higher $N$ gives the same results as low $N$ is
%general: in fact, the results at the order $N-1$ are a subclass
%of the order $N$.

\subsubsection{The sextic radial oscillator}

As we indicated above, this case corresponds to the parameter $a=0$.
From physical reason, we have to consider also $V_{-2}=0$. This is in
fact included in the constraint equations of the general case above, see
Eqs.~(\ref{Eq57}) and~(\ref{Eq63}). To simplify the discussion, we choose,
following Ushveridze, $b=1, V_2=0$.
As we have seen previously, the case $N=0$ and $N=1$ are
degenerate. We obtain
\beqn
\lambda_0&\equiv&\lambda_1=-\frac{V_1+3\sqrt{V_3}}{4\sqrt{V_3}}\label{Eq66},\\
E_0&\equiv&E_1=0,\label{Eq67}\\
\psi_0(x)&\equiv&\psi_1(x)=c_0\exp\left(-\frac{\sqrt{V_3}}{4}x^4
\right)
(x^2)^{-\frac{3}{4}-\frac{V_1}{4\sqrt{V_3}}},
\label{Eq68}\\
V_{-1}&=&\frac{\left(V_1+3\sqrt{V_3}\right)
\left(V_1+5\sqrt{V_3}\right)}{4V_3}\label{Eq69}.
\eeqn
The last equation is taken from~(\ref{Eq56}) or~(\ref{Eq62}) with
the given $a,b,V_2$ of this Section.
It is clear that the parametrization~(\ref{Eq47}) with $n=0$ satisfies it.

Let us now look  the case $N=2$.
We then have
\beqn
\lambda_2&=&\lambda_0-1,\label{Eq70}\\
E_2&\equiv&-4\sqrt{V_3}\frac{c_0^{(2)}}{c_2^{(2)}}
=-\frac{4\epsilon}{(V_3)^{1/4}}\sqrt{-\frac{1}{2}V_1
\sqrt{V_3}-3V_3},\label{Eq71}\\
\psi_2(x)&=&c_2\exp\left(-\frac{\sqrt{V_3}}{4}x^4\right)
(x^2)^{-\frac{7}{4}-\frac{V_1}{4\sqrt{V_3}}} \left[
1+\frac{\epsilon x^2(V_3)^{3/4}}{\sqrt{
-\frac{1}{2}V_1\sqrt{V_3}-3V_3}} \right],\hspace{1cm}
\label{Eq72}\\
V_{-1}&=&\frac{\left(V_1+7\sqrt{V_3}\right)
\left(V_1+9\sqrt{V_3}\right)}{4V_3}\label{Eq73},
\eeqn
where $\epsilon=\pm1$.
We see that the last equation is satisfied by~(\ref{Eq47}) with $n=1$.
For this constraint, two solutions are possible: A ground
state with $\epsilon=1$ and a first excited state with $\epsilon=-1$.
Note that the argument under the square root is non-negative.
This is seen easily with the parametrization~(\ref{Eq47}) with $n=1$.
This can also be seen from the requirement that the  wavefunctions
must vanish
at the origin, which implies $-7/4-V_1/(4\sqrt{V_3})>0$.
This is enough to ensure a positive argument of the square root.

The case $N=3$ leads to the same results as $N=2$. For $N=4$, we get
the new results
\beqn
\lambda_4&=&\lambda_0-2,\label{Eq74}\\
E_4&=&-4\sqrt{V_3}\frac{c_2^{(4)}}{c_4^{(4)}},\label{Eq75}\\
\psi_4(x)&=&c_0\exp\left(-\frac{\sqrt{V_3}}{4}x^4\right)
(x^2)^{-\frac{11}{4}-\frac{V_1}{4\sqrt{V_3}}}
\left[
c_0^{(4)}+c_2^{(4)}x^2+c_4^{(4)}x^4
\right],
\label{Eq76}\\
V_{-1}&=&\frac{\left(V_1+11\sqrt{V_3}\right)
\left(V_1+13\sqrt{V_3}\right)}{4V_3}\label{Eq77},
\eeqn
with the three possibilities
\beqn
&&c_2^{(4)}=0,c_4^{(4)}=c_0^{(4)}\frac{2V_3}{V_1+8\sqrt{V_3}},
\label{Eq78}\\
&&c_2^{(4)}=c_0^{(4)}2\sqrt{2}\epsilon
\frac{\sqrt{(-V_1-9\sqrt{V_3})V_3}}{|V_1+10\sqrt{V_3}|},
c_4^{(4)}=-2c_0^{(4)}\frac{2V_3}{V_1+10\sqrt{V_3}},\label{Eq79}
\eeqn
with $\epsilon=\pm1$.
This implies that we have three solutions corresponding to the
constraint~(\ref{Eq77}), which is satisfied with the
parametrization~(\ref{Eq47}) with $n=2$. These three solutions are
the ground state, obtained with~(\ref{Eq79}) and $\epsilon=1$,
the first excited
state, obtained with~(\ref{Eq78}), and the second
excited state, obtained with~(\ref{Eq79}) and $\epsilon=-1$.
The arguments under the square root are non-negative. The argument follows
the same lines as in the case $n=1$.

The case $N=5$ is again identical to $N=4$.

Using our algorithm, it is easy to see, and we have already mentioned it,
that $\lambda_N=\lambda_0-N/2$, or, because odd values of $N$ give
identical results to the even values, $\lambda_{2n}=\lambda_0-n$
(see Eqs.~(\ref{Eq70}) and~(\ref{Eq74})). This is true
 for arbitrary $V_1$ and $V_3$.
From Eq.~(\ref{Eq66}), a vanishing wavefunction at the origin
 implies $-V_1/\sqrt{V_3}>4n+3$.
This requires a negative $V_1$, as was also seen in the cases $n=0,1,2$ above.
If we now choose $V_1$ and $V_3$ from~(\ref{Eq47}),
the condition is translated to $s>1/4$, i.e., it becomes independant of $n$,
and the exponent of $x^2$ is just $s-1/4$, i.e., with the $n$-dependent
parametrization of $V_1$, we have $\lambda_{2n}=\lambda_0=s-1/4$.

From~(\ref{Eq69}), (\ref{Eq73}) and~(\ref{Eq77}), we can also find
the condition that has to be satisfied by $V_{-1}$:
\be
V_{-1}=\frac{\left[V_1+(4n+3)\sqrt{V_3}\right]
\left[V_1+(4n+5)\sqrt{V_3}\right]}{4V_3}.
\label{Eq80}
\ee
With $V_1$ and $V_3$ from~(\ref{Eq47}), we recover the corresponding $V_{-1}$
given by Ushveridze.

Finally, the relation between the energy and the coeffcients $c_m^{(n)}$ is
$E_{2n}=-4\sqrt{V_3}c_{n-2}^{(n)}/c_{n}^{(n)}$, i.e., only the last two
coefficients of the Taylor expansion of the wavefunction are required.
This is clearly true when considering Eqs.~(\ref{Eq67}), (\ref{Eq71})
and~(\ref{Eq75}). This is also true to higher order $n$.

It is easy, although tedious, to go to higher $n$. For concision, 
the results are not written here.

This Section has described the sextic radial oscillator. If we suppress
the  centrifugial barrier, the system is defined on the whole real
line: The potential corresponds to the sextic oscillator.
This is the subject of
the next Section.

\subsubsection{The sextic oscillator}
As in the previous Section, we restrict the analysis to $V_{2}=0$.
We can solve the problem using the results of the sextic radial oscillator,
implementing the constraint $V_{-1}=0$ by taking $s=1/4$ or $s=3/4$, i.e.,
$s=p/2+1/4$, with $p$, the parity, being equal to 0 or 1.
The parametrization~(\ref{Eq47}) leads to~(\ref{Eq64}).
Compared to the radial case, we now deal with odd or even eigenfunctions.
This is due to the fact that the potential is real and symmetric around
the origin.
The case $s=1/4$ leads to a vanishing $\lambda_N$, hence deals with
the even eigenfunctions. It corresponds to $p=0$. The case $s=3/4$,
corresponding to $p=1$, implies $\lambda_N=1/2$. Starting with the
sextic radial oscillator, defined on the positive
half line, this leads to a prefactor
$f(x)^{\lambda_N}=x$. Continuing the solution on the negative half line,
we have finally generated the odd solutions.

Since the wavefunctions correspond either to $s=1/4$ or $s=3/4$, i.e., to
two different potentials, the solutions fall into two different
 classes:
For a given sextic oscillator, it is not possible to generate simultaneously
both odd and even solutions.

\subsubsection{An exactly solvable case}

Even with $V_{-1}$ and $V_{-2}$ nonvanishing, the generalized sextic
oscillator can be reduced to an exactly solvable model: Fixing
the parameters in such a way to have $V(x)$ given by Eq.~(\ref{Eq65}),
all the solutions of the \Sch equation can be found.

Using our algorithm (\ref{Eq16}), we can see that, with $f(x)=1+x^2, h(x)=x$,
we get $\lambda_N=-1, g(x)=\exp(-x^2/4)$. The general shape of the solution
is
\be
\psi_n(x)=\frac{c_n}{1+x^2}\exp\left(-\frac{x^2}{4}\right) P_n(x^2)
\label{Eq81}
\ee
with $c_n$ a normalization coefficient
and $P_n(x^2)$ a polynomial of $x^2$ of order $n$.
The first few values are:
$P_0=1, P_1=x(3+x^2), P_2=-1+2x^2+x^4, P_3=x(-5+x^4)$. We see that a factor
$x^p$ can be factored, with $p=0$ or $p=1$, leading again, as expected,
to odd and even eigenfunctions. Our approach also shows that the
energies are related one to each other by $E_{n+1}=E_n+1$, i.e., they
are equally spaced like those of the harmonic oscillators.

In fact, using an irreducible  second order Darboux
transformation, we can relate the potential~(\ref{Eq65}) to the
potential of the harmonic oscillator. This implies that the
eigenvalues of the potential of this Section are the same (up to a
shift) as the ones of the harmonic oscillator, excepted that we
lose its two first excited levels. The Darboux transformation
also relates the corresponding eigenfunctions:
\beqn
P_0&=&1,\label{Eq82}\\
P_{n+1}&=&x(3+x^2)He_n(x)-(1+x^2)\frac{dHe_n(x)}{dx},\label{Eq83}
\eeqn
with $He_n(x)=2^{-n/2}H_n(x/\sqrt{2})$ where $H_n(x)$ is the Hermite
polynomial of order $n$.

It is trivial to check that the results coming from our approach
satisfy this relation, and that our eigenvalues are the ones
expected from the Darboux transformation. We note that since the
Darboux transformation is constructed from the oscillator
wavefunctions, the coefficients $V_{-1}, V_{-2}$ depend on the
frequency $\omega$. In the case we have investigated here, the
values $V_{-1}=4, V_{-2}=-8$ correspond to the particular choice
$V_1=1/4$, i.e., $\omega^2=1/4$.

\subsubsection{The Kuliy-Tkachuk potential}

We now turn to another class of problems: Those for which the number
of solutions is fixed from the start. As an example, we choose
the Kuliy-Tkachuk potential \cite{ref7}, for which
three solutions can be found.  It is a particular case of
the  potential~(\ref{Eq46}), but with $V_2=V_3=0$.
Before using the Kuliy-Tkachuk parameters given in Eq.~(\ref{Eq48}),
we first start with
$V_1, V_0, V_{-1}, V_{-2}$ arbitrary. Compared to the general case of
the sextic oscillator, our algorithm allows us to see that
only one coefficient $V_i$, say $V_{-2}$, is constrained, the
other ones being arbitrary. It also allows to see that the energies
take simple values in terms of the parameters and $N$:
\be
E_N=V_0+a V_1+\sqrt{bV_1}\left[(2N+1)+4\lambda_N \right],
N=0,1,2,... \label{Eq84}
\ee
However, as $N$ is increased, the number of solution is growing
because different values of the parameters are possible. For
$N=0$  two solutions are possible: There are two possible values
for $V_{-2}$ in terms of $a,b,V_1$. This leads to two possible
values for $\lambda_0$, and then to two energies and two
wavefunctions. Note that they correspond to two different values
of $V_{-2}$, hence they do not refer to the same potential. The
situation is similar for $N=1$. For bigger $N$, we face the same
difficulty as in the case of the potential~(\ref{Eq46}): Having
taken $V_2=V_3=0$ has not simplified the solving of the problem,
i.e., for general parameters, there is only one solution per
given potential. On the other hand, fixing some of the parameters
may lead to exactly solvable problems, such as in the previous
Section, or to Q.E.S. systems with a fixed number of solutions :
The K-T choice, with three solutions, corresponds to
\be V_1=\frac{3}{4}b,
V_0=\left(\frac{9}{4}-\frac{7}{2}\sqrt{3}\right)b,
V_{-1}=2(3-\sqrt{3})b, V_{-2}=(4\sqrt{3}-6)b,a=1. \label{Eq85} \ee
With this parametrization, we find, for $N=0$:
\beqn
\lambda_0&=&\frac{\sqrt{3}}{1+\sqrt{3}},\label{Eq86} \\
E_0&=&0,\label{Eq87} \\
\psi_0(x)&=&c_0\exp\left( -\frac{\sqrt{3}}{4}bx^2\right)
\left(1+bx^2\right)^{\frac{\sqrt{3}}{1+\sqrt{3}}}, \label{Eq88}
\eeqn
while, for $N=1$, we have
\beqn
\lambda_1&=&\frac{\sqrt{3}}{3+\sqrt{3}},\label{Eq89} \\
E_1&=&3(2-\sqrt{3})b,\label{Eq90} \\
\psi_1(x)&=&c_1\exp\left( -\frac{\sqrt{3}}{4}bx^2\right)
x\left(1+bx^2\right)^{\frac{\sqrt{3}}{3+\sqrt{3}}}.\label{Eq91}
\eeqn
The case $N=2$ leads to the same solution as the previous cases
$N=0,1$, as well as to the new solution:
\beqn
\lambda_2&=&\frac{1}{2}(\sqrt{3}-1),\label{Eq92} \\
E_2&=&2(3-\sqrt{3})b,\label{Eq93} \\
\psi_2(x)&=&c_2\exp\left( -\frac{\sqrt{3}}{4}bx^2\right)
(1-bx^2)\left(1+bx^2\right)^{\frac{1}{2}(\sqrt{3}-1)}.\label{Eq94}
\eeqn
For higher $N$, we only generate the three cases above. As stated
in the beginning of this Section, the Kuliy-Tkachuk potential
corresponding to~(\ref{Eq85}) allows to obtain exactly three
wavefunctions and the corresponding eigenvalues which also satisfy
the general relation~(\ref{Eq84}).

Up to now, we have only considered the simplest
variant~(\ref{Eq16}) of our algorithm. We now discuss briefly two
simple cases were the general approach~(\ref{Eq31})--(\ref{Eq30})
is needed. These cases are the Lam\'e equation and the screened
Coulomb potential.

\subsection{The Lam\'e potential}

The Lam\'e equation is a Schr\"odinger equation
 with a periodic potential of the Elliptic Jacobic type:
\be \left[-\frac{d^2}{dx^2} + k^2m_N(m_N+1)\sn^2(x,k)\right]\psi_N
(x) = E_N \psi_N(x), \; \label{Eq95} \ee
with $k\in[0,1]$. This equation admits analytic solution for
nonnegative integers $m_N$ \cite{arscott64}. Q.E.S. extensions of
the Lam\'e equation can be found in the literature, e.g.
\cite{khare01}. In the following, we restrict ourselves to
Eq.~(\ref{Eq95}). This can be examined using our generalized
algorithm~(\ref{Eq31})--(\ref{Eq30}). Several possibilities
exist, showing the flexibility of our approach. For instance,
we can choose:
\be h(x)=\sn(x,k)\rightarrow
h'(x)=\cn(x,k)\dn(x,k)=\pm\sqrt{[1-h^2(x)][1-k^2h^2(x)]}.
\label{Eq96} \ee
The $\pm$ sign implies to stay in a given interval, i.e. we
choose the plus or minus sign in~(\ref{Eq96}) the solution being
extended appropriately on the whole line. Another choice
corresponds to
\be
h(x)=\exp(i\phi)\rightarrow h'(x)=ih(x)\sqrt{1+\frac{k^2}{4}
\left[h(x)-\frac{1}{h(x)}\right]^2} \label{Eq97}, \ee
with $\phi$ the amplitude of $x$.
A third choice, which is the one we shall take to illustrate our
algorithm, is
\beqn
h(x)&=&\frac{\sn(x,k)}{\cn(x,k)}, \label{Eq98}\\
\sn^2(x,k)&=&\frac{h^2(x)}{1+h^2(x)}, \label{Eq99}\\
h'(x)&=&\sqrt{1+h^2(x)(2-k^2)+h^4(x)(1-k^2)}, \label{Eq100} \eeqn
to which corresponds $f(x)=1/\sn^2(x,k)=1+1/h^2(x)$. Hence, we have
$\Vmax=0,\Vmin=2,V_{-2}=V_0=0,
V_{-1}=k^2m_N(m_N+1),f_l^0=\delta_{l,0}+\delta_{l,-2},\sth_l=\delta_{l,0},\dth=\delta_{l,0}
+(2-k^2)\delta_{l,2}+(1-k^2)\delta_{l,4}$.

For $N=0$, our algorithm gives immediately $k=0$, or $k=1$ and/or
$g(x)=1$. In fact, the two cases $k=0$ and $k=1$ are included as
special case of the full $k$ problem, so that we can take
immediately $g(x)=1$. For a compatibility with the remaining
equations, our program shows that we need either $m=0,
\lambda_0=0, E=0$, or $m=1, \lambda_0=-1/2, E=1+k^2$. The
corresponding eigenfunctions are either a constant, or
$\sn(x,k)$, respectively. In fact, because $\lambda_0=-1/2$, we
see that the solution is $\psi(x)=[f(x)]^{-1/2}=|\sn(x,k)|$.
However such a solution is not differentiable everywhere, and
then is not a solution of the original \Sch equation. This
implies that we need to consider the problem only on half a
period of the function $\sn(x,k)$, thus allowing to remove the
absolute operation, and to continue the solution on the whole
real line, thus giving rise to $\psi(x)=\sn(x,k)$. This need of
studying the problem on a given interval was also present in the
choice~(\ref{Eq96}). The need was seen from the beginning because
of the $\pm$ sign at the level of $h'(x)$.

Putting the results together, we see that $N=0$ leads to
\beqn
m_0&=&0,\label{Eq101}\\
\lambda_0&=&0,\label{Eq102}\\
E_0&=&0,\label{Eq103}\\
\psi_0(x)&=&c_0,\label{Eq104}
\eeqn
or
\beqn
m_0&=&1,\label{Eq105}\\
\lambda_0&=&-\frac{1}{2},\label{Eq106}\\
E_0&=&1+k^2,\label{Eq107}\\
\psi_0(x)&=&c_0\sn(x,k).\label{Eq108}
\eeqn

For $N=1$, several other new solutions are found:
\beqn
m_1&=&1,\label{Eq121}\\
\lambda_1&=&-\frac{3}{2},\label{Eq122}\\
E_1&=&1,\label{Eq123}\\
\psi_1(x)&=&c_1\cn(x,k),\label{Eq124}
\eeqn
or
\beqn
m_1&=&2,\label{Eq117}\\
\lambda_1&=&-1,\label{Eq118}\\
E_1&=&4+k^2,\label{Eq119}\\
\psi_1(x)&=&c_1\cn(x,k)\sn(x,k),\label{Eq120} \eeqn
or
\beqn
m_1&=&2,\label{Eq109}\\
\lambda_1&=&-1,\label{Eq110}\\
E_1&=&2\left(1+k^2\pm\sqrt{1-k^2+k^4}\right),\label{Eq111}\\
\psi_0(x)&=&c_1\left[(1-k^2)\sn^2(x,k)+
\left(k^2\mp\sqrt{1-k^2+k^4}\right)\cn^2(x,k)
\right],\label{Eq112}
\eeqn
or
\beqn
m_1&=&3,\label{Eq113}\\
\lambda_1&=&-\frac{3}{2},\label{Eq114}\\
E_1&=&\left(5+5k^2\pm2\sqrt{4-7k^2+4k^4}\right),\label{Eq115}\\
\psi_1(x)&=&c_1\sn(x,k)\left[ \sn^2(x,k)(k^2-1)-\left(
2k^2-1\mp\sqrt{4-7k^2+4k^4} \right) \right].
\label{Eq116}
\eeqn
The solutions we got for $N=0$ were also obtained.

The analysis can of course be continued with higher values of $N$.
What we observe is the expected set \cite{arscott64} of eight
different types of solutions modulated by polynomials in $\sn$.

\subsection{The screened Coulomb potential}

The screened Coulomb potential \cite{znojil83} is also a nice
example of the working of our method. It was shown in that
reference that, under a given relation between the coefficients
of the potential $V(x)=F/x^2+G/x+H/(x+z^2)$, the solutions are
polynomials. We shall verify this fact using our formalism. We
first take $f(x)=(x+z^2)$ and $h(x)=x$ with
$V_{k}=\delta_{k,0}+\delta_{k,-1}$ and $\vf_{l,k}=0$ but for
$\vf_{l,0}=F\delta_{l,-2}+G\delta_{l,-1}$ and
$\vf_{l,-1}=H\delta_{l,0}$. We note that the coefficient $F$ has
to be $F>-1/4$. This comes from the physical requirement of
having, with the parametrization $F=L(L+1)+F_1$ and $L$ a
nonvanishing integer (angular momentum), $F_1>-(L+1/2)^2$.

Starting our algorithm with $N=0$, we obtain straightforwardly
that one of the coefficients of the potential must be expressed
in terms of the others, e.g, G:
\be G=\frac{4F+\left(1+\sqrt{1+4F}\right)
\left(4+Hz^2\right)}{2z^2}. \label{Eq125}\ee
Then, the solution can be expressed as
\beqn
\lambda_0&=&1, \label{Eq126}\\
E_0&=&-\frac{4F+2\left(1+\sqrt{1+4F}\right)
\left(1+Hz^2\right)+\left(Hz^2\right)^2}{4z^4}, \label{Eq127}\\
\psi_0(x)&=&\exp\left[\frac{x}{2z^2}
\left(1+\sqrt{1+4F}+Hz^2\right) \right]
x^{\left(1+\sqrt{1+4F}\right)/2}\left(x+z^2\right),
\label{Eq128}\eeqn
with a positive argument of the square root from the condition
$F>1/4$. Another solution is possible but we have
rejected it because it is unbounded at the origin. We also
observe the following relation between the energy, and the
different coefficients of the potential:
\be \frac{\left(G+H\right)^2}{-E_0}= \left(
3+\sqrt{1+4F}\right)^2. \label{Eq129} \ee
In fact, the vanishing of the wavefunction at infinity requires
$Hz^2<1+\sqrt{1+4F}$, see~(\ref{Eq128}). Because of this, it is
easy to verify that $(G+H)<0$. Hence, Eq.~(\ref{Eq129}) can be
replaced by
\be \frac{\left(G+H\right)}{\sqrt{-E_0}}= -\left(
3+\sqrt{1+4F}\right), \label{Eq130} \ee
a relation which was also obtained in \cite{znojil83}, with the
index $M=1$ in that reference.

The analysis can of course be done for higher $N$. The case $N=1$
is however already highly involved and will not be presented
here. We just quote that, as in \cite{znojil83}, the number 3 in
Eq.~(\ref{Eq130}) has to be replaced by $2M+1$, with $M\ge1$.

\section{A new Q.E.S. potential}
\label{Section5}

Using a Darboux transformation  on the following sextic
radial oscillator
\be
V(x)=\frac{1}{4}x^6-5x^2+\frac{35}{4x^2},
\label{Eq131}
\ee
we can show \cite{dsv} that the wave function
\be
\psi(x)=\frac{\exp\left(
-x^4/8
\right)x^{3/2}\left(6+x^4\right)}{x^8+4x^4+20}
\label{Eq132}
\ee
is a solution of vanishing energy of the \Sch equation with the potential
\be
V(x)=\frac{1}{4}x^6+x^2+\frac{3}{4x^2}
+\frac{16x^2\left(x^4-6\right)}{x^8+4x^4+20}
-\frac{2048x^6}{\left(x^8+4x^4+20\right)^2}.
\label{Eq133}
\ee
We show now this proposition using our algorithm, noting
 first that the potential obeys the parametrization 
$f(x)=x^8+4x^4+20$ and $h(x)=x$. Moreover,
$V_{k}=\delta_{k,0}+\delta_{k,-1}+\delta_{k,-2}$ and $\vf_{l,k}=0$ but for
$\vf_{l,0}=3\delta_{l,-2}/4+\delta_{l,2}+\delta_{l,6}/4$,
$\vf_{l,-1}=-96\delta_{l,2}+16\delta_{l,6}$ and $\vf_{l,-2}
=-2048\delta_{l,6}$. 

The cases $N=0$ and $N=1$ do not lead to a solution. For $N=2$,
all the constraints are fulfilled provided that
$\tg_l^1=1/8\delta_{l,4},\tg_{\log}^1=-7/2$, together with
$c_{-1}^{(2)}=c_{1}^{(2)}=c_{0}^{(2)}=0, 
c_{-2}^{(2)}=1/4(17-2\tg_{\log}^1)c_{2}^{(2)}=6c_{2}^{(2)}$,  
 and $E_2=0,
\lambda_2=1/8(\tg_{\log}^1-9/2)=-1$. The wavefunction is
then
\be
\psi_2(x)=c_2^{(2)}
\exp\left(
-\frac{x^4}{8}
\right)x^{7/2}\left(\frac{6}{x^2}+x^2\right)\left(x^8+4x^4+20\right)^{-1},
\label{Eq134}
\ee
which is nothing else than~(\ref{Eq132}).

%\section{Group theory}
%\label{Section5bis}

\section{Conclusions}
\label{Section6}

We have proposed a general method for determining the analytic eigenfunctions
(and associated eigenvalues) of a given (quasi) exactly solvable \Sch
Hamiltonian. This method consists in a general algorithm replacing
the differential \Sch equation by a finite set of algebraic equations,
these ones being treated through a Mathematica implementation.
Our general algorithm might be considered as relatively heavy from
a purely technical point of view but it actually works straightforwardly.
Moreover, for the majority of analytically solvable \Sch equations, 
it can be replaced by a simpler version as presented in 
Section~\ref{Section2}. In all cases, it gives rise in a very systematic way 
to the solutions of the \Sch equations and also precises the conditions
under which these equations are solvable. This systematization is 
particularly useful at the level of quasi exactly solvable \Sch equations
for which one knew before our approach that some of them are
relevant of $sl(2,R)$ and the others not. These two different points of view
have been unified through our formalism. We also assert that new
 quasi exactly solvable potentials (see Section~\ref{Section5}) can be
handled within our algorithm due to the fact that the group
theoretical approach is not needed anymore.

We are also convinced that it is possible to extend our algorithm to
the relativistic context, studying more particularly the  quasi exactly 
solvable Dirac equations. Only a few studies, see for instance 
Ref.~\cite{ref17}, deal with these equations because, by opposition to some
of the \Sch ones, they are not in general subtended by a specific Lie
(super) algebra. Our approach avoiding this group theoretical aspect should
be of interest in the analysis of such Dirac equations. We plan to come back
on these developments in the near future.

\section*{Acknowledgments}
We thank Prof. Y. Brihaye for numerous useful discussions at various stages of
this work and for interesting clarifications.
We also thank Prof. B. Samsonov for several hints on Darboux transformations, 
and Prof. K. Penson for suggesting us the example of the screened Coulomb
potential.
The work of N. D. and B. VdB
was supported by the Institut Interuniversitaire des
Sciences Nucl\'eaires de Belgique.

%\newpage


\begin{thebibliography}{99}
\bibitem{ref1} M. Razavy, Am. J. Phys. {\bf 48} (1980) 285; Phys. Lett.
 {\bf A82} (1981) 7.
\bibitem{ref2} A.V. Turbiner and A.G. Ushveridze, Phys.Lett. {\bf
            A126} (1987) 181.
\bibitem{ref3} A.V. Turbiner, Comm. Math. Phys. {\bf 118} (1988) 467.
\bibitem{ref4} A.G. Ushveridze, {\em Quasi-Exactly Solvable Models in Quantum
Mechanics}, IOP Publishing Ltd (1994).
\bibitem{ref5} V.M. Tkachuk, Phys.Lett. {\bf A245} (1998) 177.
\bibitem{ref6} S.N. Dolya and O.B. Zaslavskii, J. Phys. {\bf A34} (2001) 1981.
\bibitem{ref7} T.V. Kuliy and V.M. Tkachuk, J. Phys. {\bf A32} (1999) 2157.
\bibitem{ref8} Y. Brihaye, N. Debergh and J. Ndimubandi,
%{\em On a Lie algebraic approach of quasi-exactly solvable potentials
%with two known eigenstates}, quant-ph/0104009.
Mod. Phys. Let. {\bf A16} (2001) 1243.
\bibitem{ref9} E. Witten, Nucl. Phys. {\bf B188} (1981) 513.
\bibitem{yves95}
Y. Brihaye and P. Kosinski, J. Math. Phys. {\bf 36} (1995) 4340.
\bibitem{bw01} C. M. Bender and Q. Wang, {\em A class of exactly-solvable
eigenvalue problems}, math-ph/0109007.
\bibitem{khare01} A. Khare,
%{\em A QES band-structure problem in one
%dimension}, quant-ph/0105030.
Phys. Lett. {\bf A288} (2001) 69.
\bibitem{ref10}
L. Skala, J. Cizek, J. Dvorak and V. Spirko, Phys. Rev. {\bf A53} (1996) 2009.
\bibitem{dsv}
N. Debergh, B.F. Samsonov and B. Van den Bossche,
{\em Darboux transformations for quasi-exaclty solvable Hamiltonians},
in preparation.
\bibitem{arscott64}
F. M. Arscott, {\em Periodic differential equations}, Pergamon,
Oxford (1964).
\bibitem{znojil83}
M. Znojil, Phys. Lett. {\bf A94} (1983) 120.
\bibitem{ref17}
Y. Brihaye and P. Kosinski,
Mod. Phys. Lett. {\bf A13} (1998)
1445.
\end{thebibliography}
\end{document}